# TWO METHODS FOR ISO-SURFACE EXTRACTION FROM VOLUMETRIC DATA AND THEIR COMPARISON

**Vaclav Skala[i], Alex Brus[ii]**

Department of Computer Science and Engineering[iii]
University of West Bohemia, Univerzitni 8, Box 314
306 14 Plzen, Czech Republic
Skala@kiv.zcu.cz        http://iason.zcu.cz/~skala

**ABSTRACT**

There are various methods for extracting iso-surfaces from volumetric data. Marching cubes or tetrahedra or raytracing methods are mostly used. There are many specific techniques to increase speed of computation and decrease memory requirements. Although a precision of iso-surface extraction is very important, too, it is not mentioned usually. A comparison of the selected methods was made in different aspects: iso-surface extraction process time, number of triangles generated and estimation of radius, area and volume errors based on approximation of a sphere. Surprisingly, experiments proved that there is no direct relation between precision of extracted and human perception of the extracted iso-surface.

## 1. INTRODUCTION

Volume data sets are often acquired by scanning a material of interest by using Magnetic Resonance Imaging (MRI), Computed aided Tomography (CT), Positron Emission Tomography (PET) etc. Particular data acquisition devices, such as CT and MRI scanners, are good at sampling a specific characteristic of a substance. This characteristic is usually different for each kind of device used, e.g. MRI scan shows soft tissue, while CT scan shows hard material, like bones etc.

The volume data representation is not restricted to the medical applications only, but there are also practical technical problems that can be solved using this representation, i.e. this representation can be used for computational and visualization purposes. The advantage of this representation is that all volume data sets can be treated similarly even though they are generated by diverse means. The amount of processed data is based on application, but typical experimental data size are 512 x 512 x 256 voxels, 2 Bytes per voxel.

A resolution in x, y and z axes is different generally, but in many applications, especially in medical applications, the resolution in x and y axes is the same, see Fig.1.1.

It is necessary to point out that today's Intel based systems are equipped usually with high performance graphics card, memory up-to 2 GB RAM and 500-600 MHz processor Pentium III. It moves some recently considered problems away and others become more important.

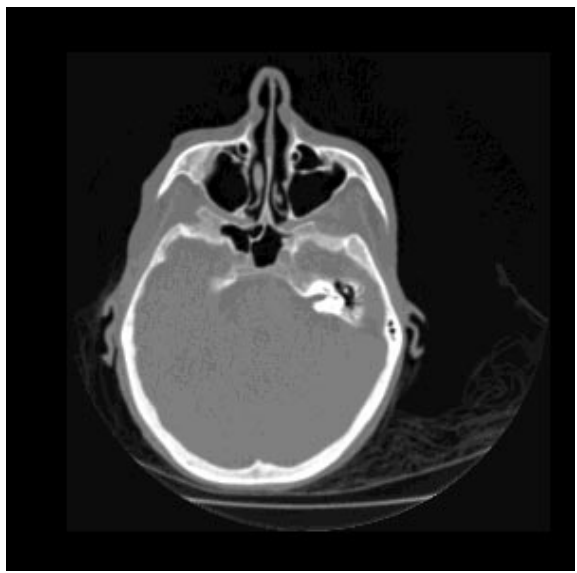

Typical CT scan for x-y plane
Figure 1.1

The volume data representation differs from the traditional computer graphics data structures in:
- the data from which images are formed are captured from real objects rather than

- synthesised by a designer or generated mathematically from a computer model,
- accurate iso-surface extraction is more important than pleasing images,
- huge number of elements are processed and generated for volumetric data sets.

Volume data visualization algorithms can be split into two main groups [Chua95a]:

- Direct Volume Rendering (**DVR**) algorithms that maps elements directly to the screen space without using geometric primitives as a temporary representation. These methods are mostly used for creating images from data sets. One fundamental disadvantage of DVR method is that the entire data set must be traversed each time the image is rendered, e.g. when the viewing angle is changed or zoom is used etc.
- Surface-Fiting (**SF**) algorithms use surface primitives such as polygons (typically triangles) or patches to represent iso-surface in volumetric data sets. The SF approach includes contour-connecting, marching cubes, marching tetrahedra, dividing cubes, and others.

Generally the SF methods are usually faster than DVR methods as SF methods traverse the volume only once to extract surfaces. After extracting the surfaces, rendering hardware and rendering methods can be used to fast rendering of selected iso-surface. It is necessary to point out that changing the SF threshold value means all the cells to be processed again to extract a iso-surface.

Table 1 shows a taxonomy of a well-known algorithms [Tod92a]:

| Volume Visualization Algorithms | | |
|---|---|---|
| **Surface-fitting** | **Direct Volume Rendering** | |
| Opaque cubes (Cuberille) Contour connecting Marching Cubes Marching Tetrahedra | Projection methods | Image-order methods |
| | V-buffer Splatting | Ray-casting Cell integration Sabella method |

Well known approaches
**Table 1**

There are several approaches to the iso-surface extraction from volumetric data. Described methods are usually compared according to speed of computation, but there are other properties that must be considered when the method is to be used in visualization pipeline, e.g. mesh property, number of generated triangles, whether an information on triangle neighbour is available (needed for mesh reduction) etc. One very important property of iso-surface extraction is the precision of the found surface or volume as it is not equivalent to the human perception of final picture "quality".

## 2. MARCHING CUBES

Marching Cubes (MC) method is based on voxel vertices classification. It is known that there are 256 possible cases but they can be reduced to 15 fundamental (enlarged vertex means that the voxel value is above a threshold), see Fig.2.1.

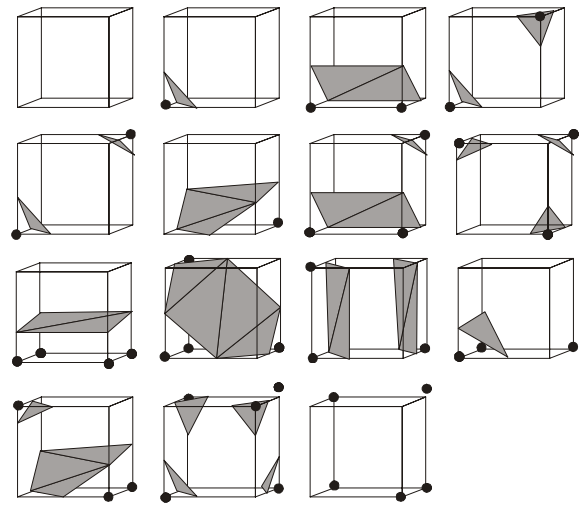

Fundamental cases in MC method
Figure 2.1.

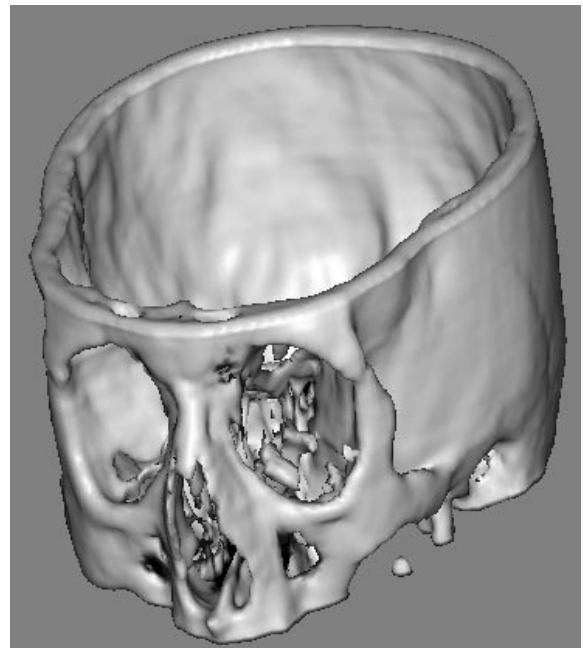

MC method - computational time 0.65s
93 395 vertices, 185 535 polygons
Figure 2.2

The advantage of this method is that it generates generally less triangles than Marching Tetrahedra method, see Fig.2.2. and Fig. 3.1. for time and memory comparisons.

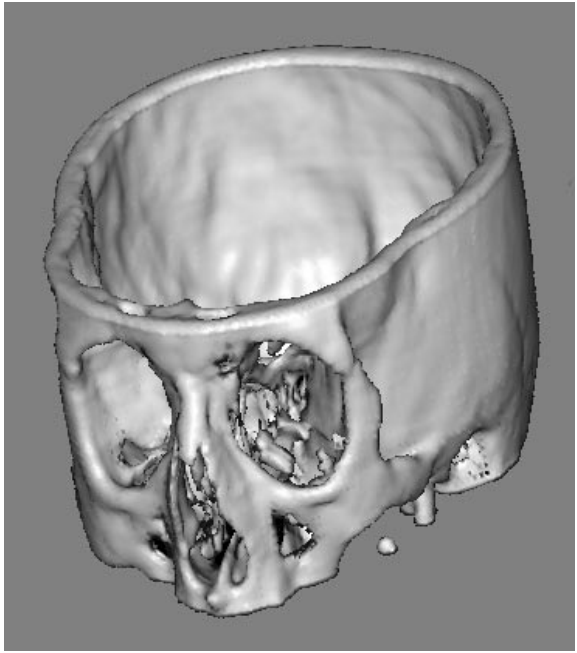

MT method, 5 tetrahedra, computational time 1.19s
230 015 vertices, 460 068 polygons
Figure 3.1

## 3. MARCHING TETRAHEDRA

Marching Tetrahedra (MT) method is based on a different approach based on decomposition of the voxel space to tetrahedron space, that gives generally only two fundamental cases as the tetrahedron can be intersected by iso-surface only in 4 points. It means that the iso-surface can be represented by two triangles, generally. The advantage of this approach is that it generates final image principally without "holes" that might happen in MC algorithms.

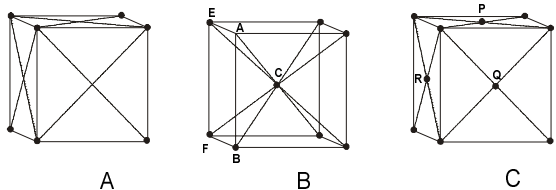

Figure 3.2

Generally three cubic lattices are used, see fig.3.2, and following fundamental cases can be distinguished: cubic primitive (A), body-centred cubic (B), face-centred cubic (C). It is well known that five tetrahedrons is the smallest number how cube can be partitioned. There are several conform cube decomposition, for details see [Kol94a].

There are two fundamental decomposition to:
- 5 "corner" tetrahedra scheme gives 4 identical tetrahedra and one different - centred. It is twice big as the corner terahedron,
- 6 tetrahedra scheme gives the advantage of common faces for tetrahedra of neighbour voxels, see fig.3.3. This is very important feature for iso-surface generation.

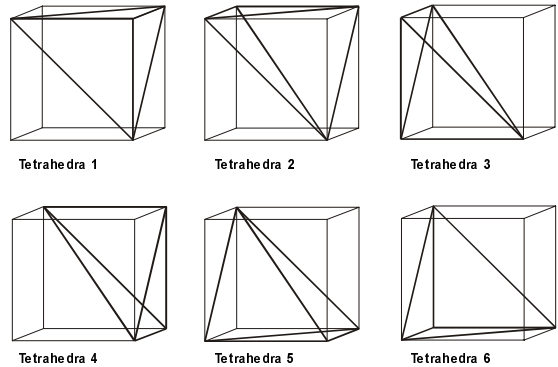

Possible decomposition to 6 tetrahedra
Figure 3.3

If we use body-centred and face-centred cubic lattices, we can get 6, 12, 24, 36, 48 tetrahedrons per voxel. This approach generates more triangles, so it has very high computational and memory requirements. The result of experiments proved that error of iso-surface extraction is not related to the picture perception, see Fig.3.6, for details [Bax97a]. It can be seen that the 24 tetrahedron schemes gives better result in some details then 6 tetrahedra scheme, but computational time and memory requirements are unacceptable.

There are methods based on body-centred cubic lattice that construct tetrahedrons using two neighbouring voxels [Chan98a], see Fig.3.5. It can be expected that different decomposition schemes give different properties of the final iso-surface.

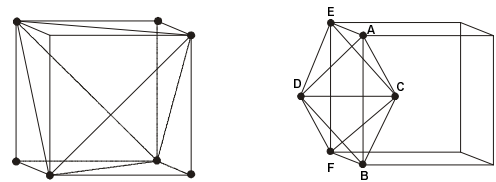

Classical 5/6 tetrahedra scheme
Figure 3.4

New 12 tetrahedra per voxel scheme
Figure 3.5

This new scheme has an advantage that all tetrahedra are the same (edges are $a$, $a\sqrt{3}$, $a$, $a\sqrt{3}$ if $2a$ is the length of the voxel's edge). Nevertheless it is necessary to consider more neighbours for iso-surface extraction than in the body-centred tetrahedronal scheme.

The properties of the standard 12 - 48 tetrahedra schemes are known [Bax97a]. Therefore the classical 6 tetrahedra and new 12 tetrahedra schemes has been compared according to time, number of triangles generated, radius, area (surface) and volume errors. Also perception of image comparison has been tested.

It is necessary to point out that error measurement in some cases is not standardised or defined, so this measurement has been defined for the purpose of this comparative study.

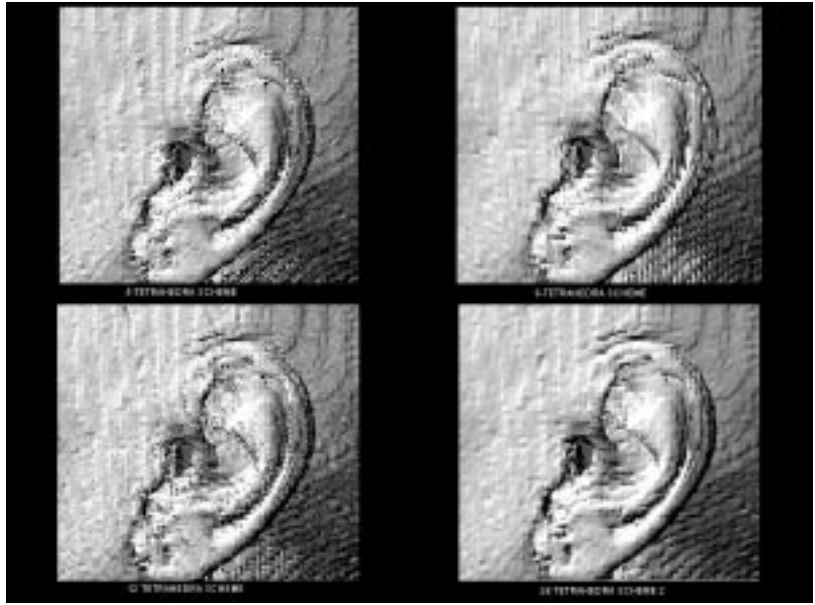

Comparison of 5, 6, 12, 24 decomposition on ear detail
Figure 3.6

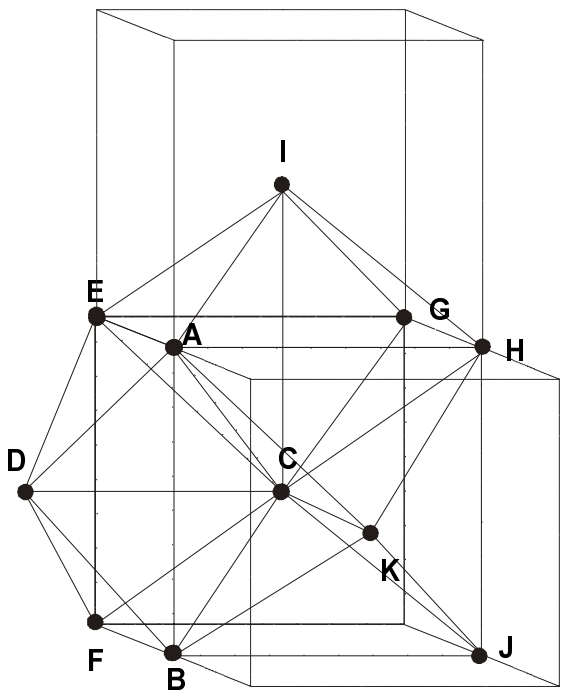

Tetrahedra that must be inspected when one voxel is evaluated
Figure 3.7

## 4. EXPERIMENTAL RESULTS

An algorithm with decomposition to 6 tetrahedra will be referenced as the First algorithm and tetrahedral decomposition of space from the centred cubic with 12 tetrahedra lattice will be referenced as the Second algorithm. In the following

- the time needed to create the iso-surface from the volumetric data is presented in section A; it is necessary to point out that no specific techniques were not employed as it is expected that those would speed up the algorithms similarly,
- a comparison of the number of triangles extracted for the iso-surface representation is presented in section B,
- sections C, D, E and F present differences between a perfect sphere and a sphere constructed using each method are measured. Various comparison have been made: radius from centre of the sphere to each vertex, radius from centre of the sphere to the centre of gravity, area of the triangles used for the creation of the iso-surface and finally volume defined by the iso-surface.

The appendix presents images from simple to complex volumetric data and visual impression that have been used to derive conclusions.

### A. CONSTRUCTION TIME

One important aspect when comparing two methods is the time for generating the iso-surface. Table 2 presents some experimental results.

| Image | | Time (seconds) | |
|---|---|---|---|
| Name | Size | First Alg. | Second Alg. |
| Syn 64 | 64x64x64 | 2 | 4 |
| Engine | 256x256x110 | 53 | 159 |
| 3dhead | 256x256x109 | 104 | 215 |
| 51Sphere | 51x100x100x100 | 124 | 432 |

Iso-surface extraction time from volumetric data
Table 2

The following features have been observed:
1. The First algorithm always spend less time than the Second one. This difference can be because of various reasons:
   - The Second algorithm must calculate the central cube positions and gradients plus the same calculations performed by the first algorithm when a cube is processed.
   - Calculations of gradients are more complex in the second algorithm (the same operations are made but the central difference method in four more positions must be evaluated).
   - The Second algorithm splits the cube to more tetrahedra so more calculation is needed.
   - The Second algorithm, during the evaluation process, have to decide, weather the cube is be processed, three central cube values are calculated by averaging eight cube vertices and this operation is performed for each cube of the model.
2. The time ratio of two methods isn't constant (sometimes it's two times higher, sometimes three times, etc.) - it depends on the data set. Nevertheless they are in relation with expectation. This variation of the ratio is because of a difference of the threshold selection; some tetrahedra are taken into account or left and this difference isn't changing in the same magnitude in both methods because the second one uses the interpolated points and this changes the ratio of those two methods.

It's possible to decrease the execution time using acceleration for the values calculated from different columns and different slices (actual implementation only use the calculated values of the previous processed cube and it's only in the row direction). With this modification the second method would benefit more than the first because operations have a higher cost.

It's important to know how much time is spent by each method without handling any tetrahedra. (constant time for explore all cubes without counting the time when the cube is handled). The way to overtake this value was using one volumetric data (3dhead with a size of 256x256x109) and execute the iso-surface process extraction with a threshold bigger than any vertex value and so any tetrahedra was handled.

The results of this execution were: the first algorithm spend 11 seconds to cross all the data volume looking for which cubes contribute in the iso-surface construction and 48 seconds for the Second algorithm. It's clear there is a different **constant time** associated to each method.

It necessary to point out that the finality of this implementation wasn't the execution speed. It was important to have a good structure and that both methods have the same method in accessing . The work has not been directed to the fastest visualization, but to the understanding what error behaviour can be expected from two fundamental approaches of iso-surface generation.

### B. NUMBER OF TRIANGLES

Form the Chart 1 and Table 3, it is clear that the second algorithm always generates more triangles than the first algorithm. The ratio of triangle generation is between 1.5 and 3. Generation of higher number of triangles causes higher memory requirements, if triangles are kept in a list, and higher execution time.

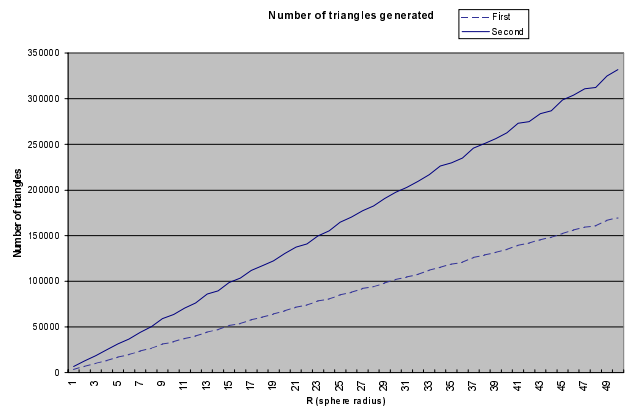

Number of triangles when construction of 51 spheres with a radius varying from 1 to 501 in steps of 10 is performed (henceforth 51Sphere).
Chart 1

This result must be one of the reasons of the difference in generation time discussed in the previous section. The cube in the second method participate in the double amount of tetrahedra than the first method and it involves more triangles.

| Image | Number of triangles | |
|---|---|---|
| | First Algorithm | Second Algorithm |
| Syn_64 | 171276 | 269184 |
| Engine | 5744340 | 8333868 |
| 3dHead | 12713532 | 17652630 |

Number of triangles used for isosurface construction of several images
Table 3

## C. RADIUS ERROR FROM EACH VERTEX

In this section there is presented the error committed when approximating a sphere, with a measure of quality the distance from the centre of the sphere to the vertexes of each triangle.
This error is calculated with the following formula:

$$\sum_{i=1}^{n} \frac{R_i - R_r}{R_r}$$

Where: n is the number of triangles multiplied by 3, $R_i$ is the radius of the sphere and $R_r$ is the distance from the centre of the sphere ($c_x, c_y, c_z$) to a vertex of a triangle ($v_x, v_y, v_z$):

$$r_i = \sqrt{(v_x - c_x)^2 + (v_y - c_y)^2 + (v_z - c_z)^2}$$

In Chart 2 it can be observed that radius error in the second algorithm is higher than in the first algorithm.

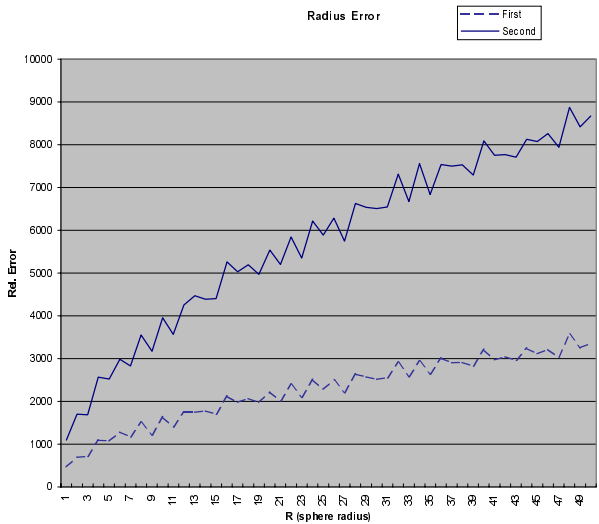

Radius Error when construction of the 51Sphere is performed.
Chart 2

In Chart 3 there are presented the radius error divided by the number of triangles generated. The first method has associated a less radius error. It means that the first algorithm is a better approximation if radius error is taken into account.

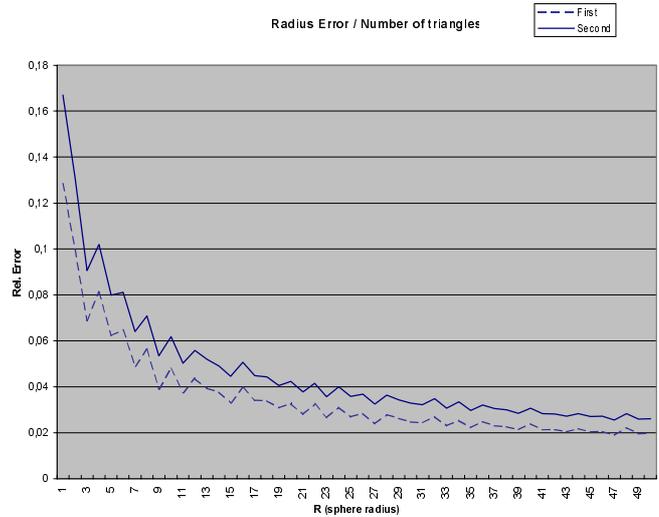

Radius Error divided by the number of triangles used when construction of the 51Sphere is performed.
Chart 3

## D. RADIUS ERROR FROM CENTRE OF GRAVITY

In this section $R_i$ is calculated from the centre of the sphere to the centre of gravity of each triangle.

Results from this approach are similar as the Radius error from each vertex.

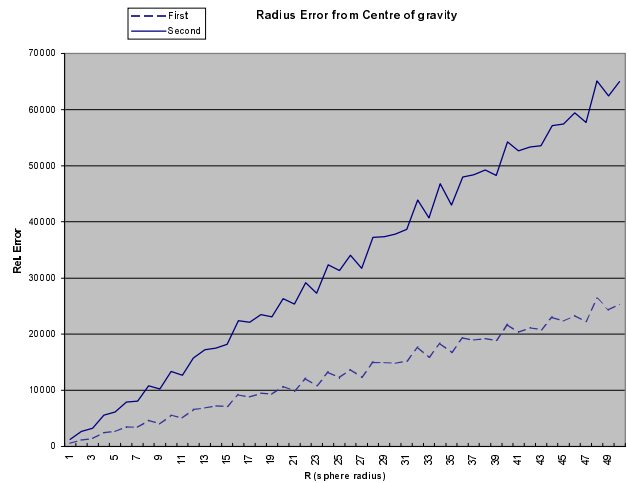

Radius Error from centre of gravity obtained when construction of the 51Sphere is performed.
Chart 4

## E. AREA ERROR

In this section there is presented comparison of triangle's area when creating the iso-surface with the theoretical value.

The magnitude of the cross product of two vectors is the area of the parallelogram they determine: If A and B are vectors, the $|A \times B|$ is the area of the parallelogram with sides A and B. Since any triangle can be viewed as half of a parallelogram, this gives an immediate method of computing the area from co-ordinates.

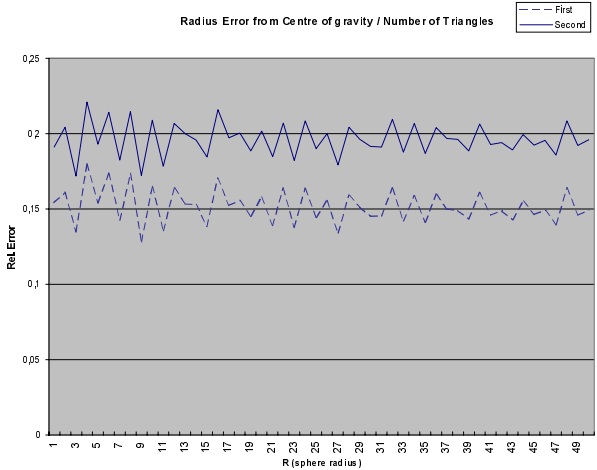

Radius Error divided by the number of triangles used for construction of the 51Sphere.
Chart 5

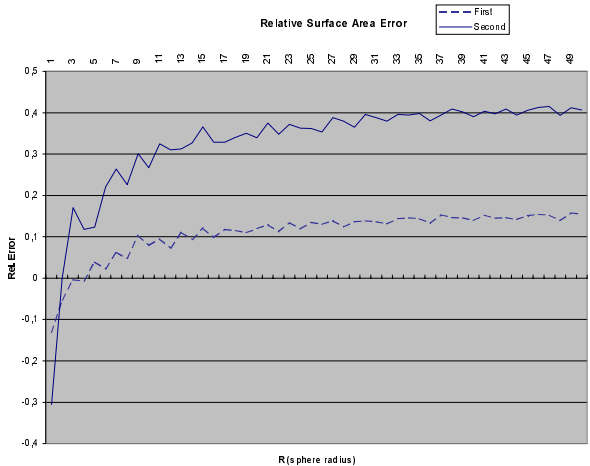

Relative surface area error of each sphere when construction of the 51Sphere is performed.
Chart 6

The theoretical area is calculated by and Relative Error by

$$E = \frac{S_d - S_r}{S_r}$$

$$S_r = 4\pi r^2$$

It can be concluded with results shown in Chart 6 that first method is a better approximation of area.

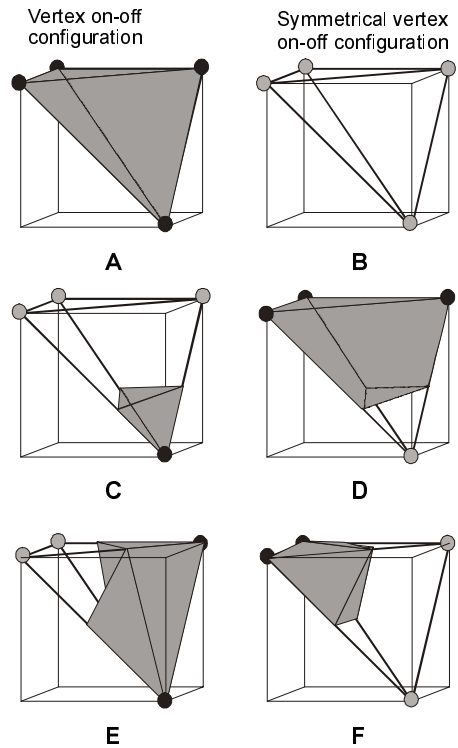

Example of volumes calculated for each configuration.
Figure 4.1

**F. VOLUME ERROR**

When calculating the surface's volume($V_d$), defined by the triangles drawn inside of the tetrahedra, the following patterns can be used:

**A Group**. All vertex of the tetrahedra have the same state:
- All vertex have the on state (all tetrahedra is inside the surface). The volume is $Vol_t$ were $Vol_t$ is the volume of the tetrahedra. Fig.4.1(A)
- All vertex have the on state off (all tetrahedra is outside the surface). The volume is zero. Fig.4.1(B)

**B Group**. One vertex $v_t$ of the tetrahedra has a different state:
- If $v_t$ has the on state (this vertex is inside of the surface and the other three are outside). The volume is $Vol_{a,b,c,t}$ were $Vol_{a,b,c,t}$ is the volume of the tetrahedra defined with the vertex $v_t$ and the three points of the triangle drawn. Fig.4.1(C)
- If $v_t$ has the off state. The volume is $Vol_t - Vol_{a,b,c,t}$. Fig 4.1(D)

**C Group**. Two vertex $v_{t1}$ and $v_{t2}$ of the tetrahedra have a different state:

- If $v_{t1}$ and $v_{t2}$ have the on state. The volume is $Vol_{a,b,c,t1} + Vol_{a,b,d,t2} + Vol_{a,b,t1,t2}$ ,see Fig.4.1(E). Fig.19 shows tetrahedra for clarification.
- If $v_{t3}$ and $v_{t4}$ have the off state. The volume is $Vol_t - (Vol_{a,b,c,t1} + Vol_{a,b,d,t2} + Vol_{a,b,t1,t2})$ Fig.4.1(F)

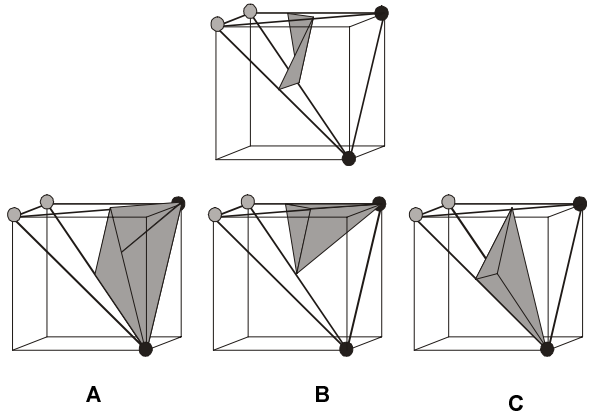

A      B      C

Black circumferences represent the on state. The above cube shows the two triangles drawn inside the second tetrahedra. The A, B and C cubes shows the three cubes used to calculate the volume.
Figure 4.2

The theoretical volume is calculated as:

$$V_r = \frac{4}{3}\pi r^3$$

The relative error is:

$$E_r = \frac{(V_d - V_r)}{V_r}$$

From Chart 7 it can be observed, the first algorithm is a better approximation of the volume than the second.

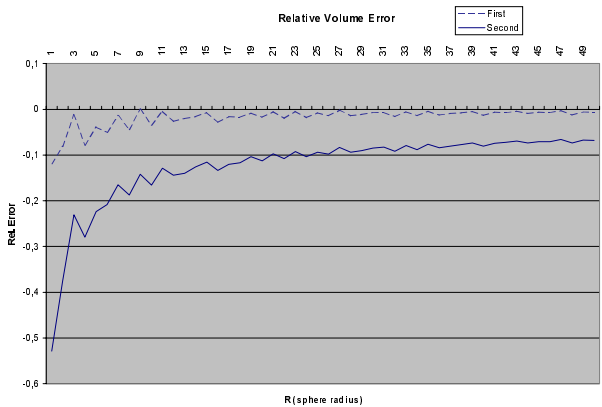

Relative volume error of each sphere when construction of the 51Sphere is performed
Chart 7

## G. IMAGE COMPARISON

When a little cube is approximated, the first algorithm gives better result, Fig.A.1. When a medium cube is drawn it can be observed that in the centre of each face the cube is drawn with the same appearance but in the each edge the second algorithm is worst, see Fig.A.2.

The reason of this result is because positions outside the cube have a zero value and positions inside have the same value as the threshold. When the averaging values of a tetrahedra with one vertex inside the object and the other outside of a long edge.

If a cube is constructed with a different threshold than the exact value, the shape of the object varies a lot and the second has an image with a similar quality of the first, see Fig.A3. Fig.A.4 illustrates two spheres and the sphere made by the second algorithm has a better shadows.

The second method is better than the first one when a complex model is generated, see Fig.A.5. It shows that images generated by the second algorithm are better: the ear and nose are more defined, the contour of the eye is better, the back of the head has more realistic relief, the mouth is better and the lips have a better shape.

## 5. CONCLUSIONS

A comparison of the Marching Tetrahedra algorithm with a decomposition in six tetrahedra and Marching Tetrahedra with a tetrahedral decomposition of the space from the cubic lattice has been presented.

When the measure of quality is generation time or number of triangles, the decomposition in 6 tetrahedras has a better results.

Drawing spheres with a different radius, distance error from each vertex of the triangles, distance error from the centre of gravity, area and volume error were studied and the experimental results proved that it is useless to implement the decomposition of the cube using the centre cubic lattice with this criteria.

Nevertheless, the visual impression is better when a complex model is generated with the second the tetrahedral decomposition of the space from the cubic lattice. This results are similar as the comparison made in [Bax97a] where five and six tetrahedra decomposition were compared with 24 and 48.

A similar explanation can be made about the better image results with the second decomposition: human eye is sensitive to edges and gradient changes of a surface.

It is expected that future work will elaborate this phenomena more in detail.

## 7. APPENDIX A

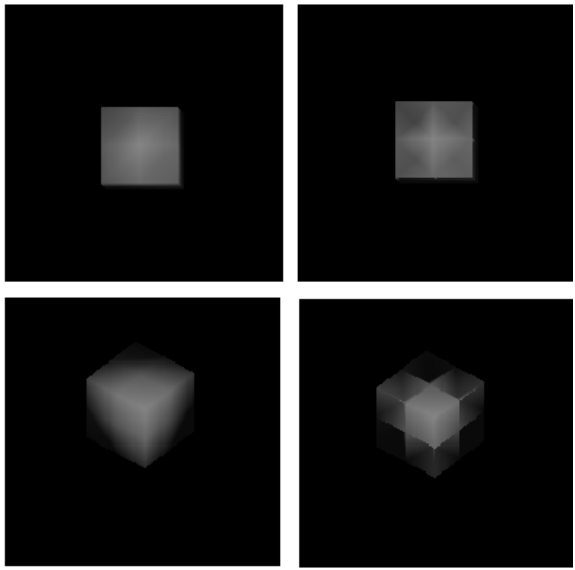

Approximation of a cube with a edge size 2. Vertex of the cube has value 100 and the others have 0. Threshold is 100. Left images are obtained from the First algorithm and right images from the Second one.

Figure A.1

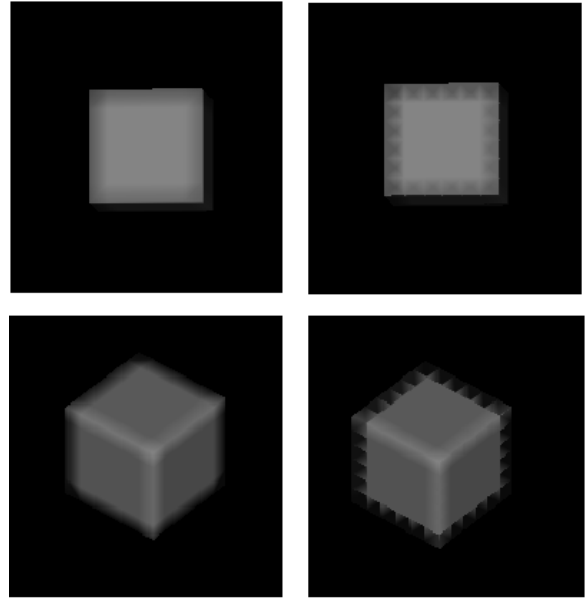

Approximation of a cube with a edge size 6. Vertex of the cube has 100 value and the others have value 0. Threshold is 100. Left images are obtained from the First algorithm and right images from the Second Algorithm.

Figure A.2

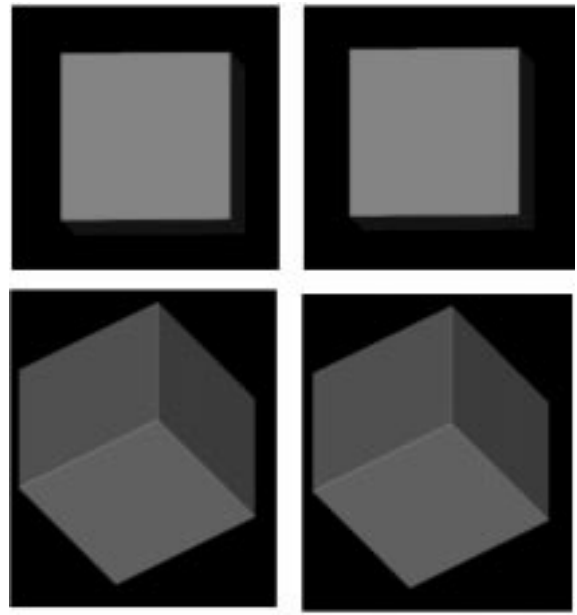

Approximation of a cube with a edge size 80. Vertex of the cube have 100 value and the others have 0. Threshold is 100. Left images are obtained from the First algorithm and right images from the Second Algorithm.

Figure A.3

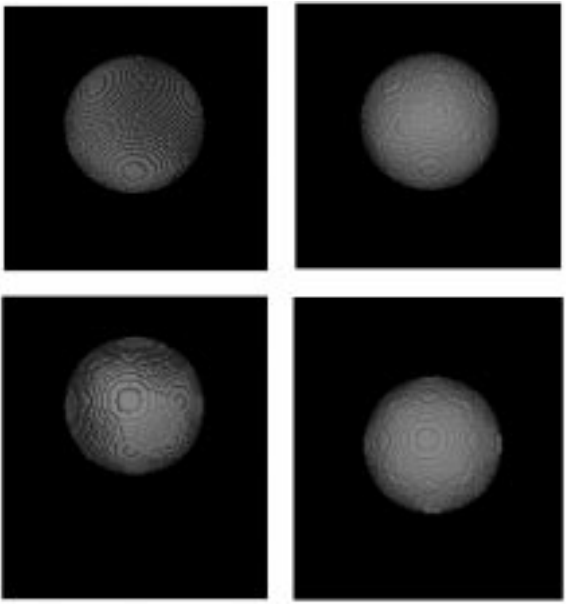

Circumference with light. Vertexes of the sphere have 100. Value and the others have 0. Sphere radius is $\sqrt{810}$. Left images are obtained from the First algorithm

Figure A.4

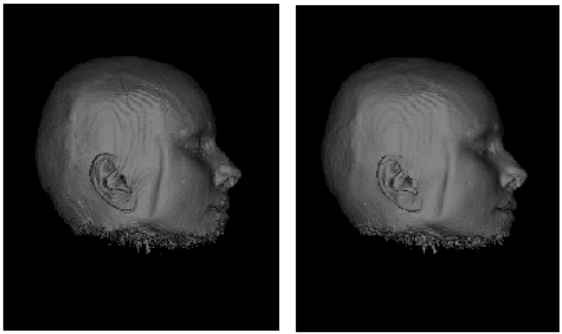

3dhead image, the left image is generated by the First algorithm, the right image by the Second one

Figure A.5

---

[i] Affiliated with Multimedia Technology Research Centre, University of Bath, BATH BA2 7AY, U.K.
[ii] University of Girona, Spain
[iii] This work was supported by the Ministry of Education of the Czech Republic - projects A2030801, VS 97155